*Article*

# Electron Beam Characterization of REBCO-Coated Conductors at Cryogenic Conditions

**Michal Haubner [1,2,*], Patrick Krkotić [1], Catarina Serafim [1,3], Valentine Petit [1], Vincent Baglin [1], Sergio Calatroni [1], Bernard Henrist [1], Artur Romanov [4], Teresa Puig [4] and Joffre Gutierrez [4]**

1 European Organization for Nuclear Research (CERN), 1211 Geneva, Switzerland
2 Department of Physics, Faculty of Mechanical Engineering, Czech Technical University in Prague, 166 07 Prague, Czech Republic
3 Department of Physics, University of Helsinki, FI-00014 Helsinki, Finland
4 Institute of Materials Science of Barcelona, CSIC, 08193 Madrid, Spain
* Correspondence: michal.haubner@cern.ch

**Abstract:** Particle accelerators with superconducting magnets operating at cryogenic temperatures use a beam screen (BS) liner that extracts heat generated by the circulating bunched charge particle beam before it can reach the magnets. The BS surface, commonly made of high–conductivity copper, provides a low impedance for beam stability reasons, low secondary electron yield (SEY) to mitigate the electron–cloud (EC) effect, and low electron–stimulated desorption yield (ESD) to limit the dynamic pressure rise due to EC. Rare–earth barium copper oxide (REBCO) high–temperature superconductors (HTSs) recently reached technical maturity, are produced as coated conductor tapes (REBCO–CCs), and will be considered for application in future colliders to decrease the BS impedance and enable operation at around 50 K, consequently relaxing the cryogenic requirements. Aside from HTS properties, industry–grade REBCO–CCs also need qualification for EC and dynamic vacuum compatibility under accelerator–like conditions. Hence, we report the SEY and ESD measured at cryogenic temperatures of 12 K under low–energy electron irradiation of 0–1.4 keV. We also verify the sample compositions and morphologies using the XPS, SEM, and EDS methods. The energy and dose dependencies of ESD are comparable to those of technical–grade metals and one sample reached $SEY_{MAX}$ = 1.2 after electron conditioning.

## 1. Introduction

High–temperature superconductors (HTSs) constitute a class of highly correlated anisotropic systems that theoretically have proven to be extremely difficult to address. Since their discovery in 1986 [1], numerous studies have been conducted to describe these materials, as well as to produce them at a scale suitable for large–scale technical applications. The general interest lies in the fact that the critical surface of HTS is significantly higher than that of low–temperature superconductors (LTSs), as visible in Figure 1a. Consequently, work gradually proceeded on technical applications, leading to large–scale HTS–based prototype devices over the years, and demonstrating the possible performance advantages over devices based on conventional LTS [2] or even normal–conducting technology.

The most promising industry-grade method available for a large–scale HTS application is by coating a thin layer of rare–earth barium copper oxide (REBCO) material on flexible metal tapes, in a complex–layered structure containing buffering metal oxide underlayers and stabilizing and protecting Cu and Ag overlayers. This technology is known as a second-generation coated conductor (CC), as depicted in Figure 1b, and is already commercially available in kilometer–lengths from several manufacturers. These second–generation REBCO–CC tapes promise high–performance properties, such as high current densities and compatibility with high magnetic field applications, and their matured technology and mass production make them an affordable and viable engineering solution.

Aside from the imminent application in high–temperature superconducting magnets, in the next–generation, particle accelerators [3,4], passive microwave devices (not generating power) will be equally strong candidates for eventual HTS–coating commercialization. A HTS can have lower microwave surface impedance compared to pure metal [5]. Subsequently, this could allow the design of high–performance RF devices, such as high–quality factor (Q–factor) resonators, narrow–band filters, and antennas with improved functioning over conventional normal–conducting devices. Additionally, such complex microwave signal transmission and signal–processing systems could be designed to be more compact and more efficient [2,6]. For instance, a potential application lies within future space–based communication systems with a more effective strategy by producing fewer and lighter high-performance satellites with the reduced weight of cavities and filters. Furthermore, the feasibility of coating curved walls with REBCO–CCs enables manufacturing high Q–factor RF cavities, such as the one developed within the relic axion detector exploratory setup (RADES) [7]. Here, the detector sensitivity scales linearly with the Q–factor of the 9 GHz resonant cavity held at 4.2 K, and REBCO is the essential technology enabling operation in a strong magnetic field of 11 T while maintaining a high Q–factor.

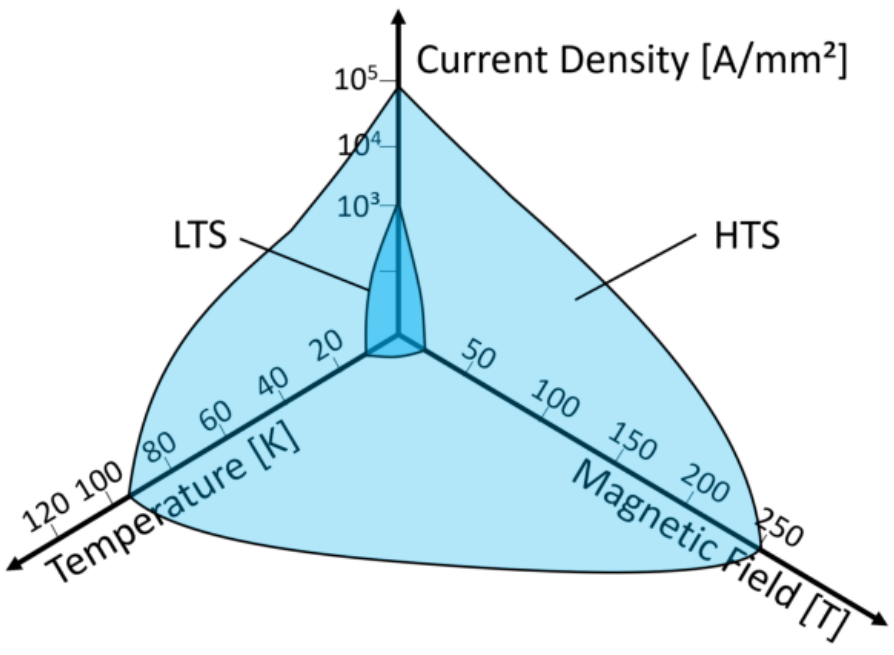

(**a**) Critical surface of LTS and HTS.

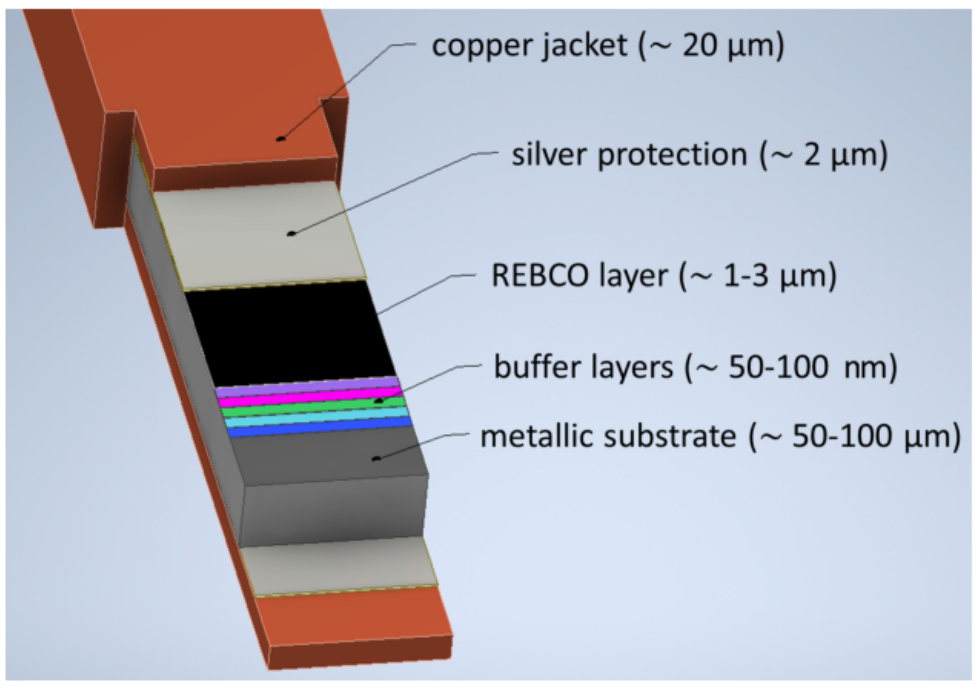

(**b**) Layers of the REBCO–coated conductor tape.

**Figure 1.** (**a**) Comparison of the critical surface phase space for LTS and HTS formed by the three interdependent characteristic parameters of a superconductor, i.e., the critical temperature, critical magnetic field, and critical current density. Superconductivity can only occur below the critical surfaces, which are unique for each material. The figure is inspired by [8]. (**b**) Structure of a complex–layered REBCO HTS–coated conductor tape. Note the substrate and buffer layers residing under the REBCO coating, the silver and copper stabilizing, and protecting overlayers.

In particle accelerators with magnets operating at cryogenic temperatures, a beam screen (BS) inserted into the cryomagnet apertures is used to intercept the heat and radiation generated by the circulating particle beam before it can reach the superconducting magnets. The BS can equally be considered an RF device due to the time–variant electric field induced by the passing bunched charged particle beam. Taking the BS in the Large Hadron Collider (LHC) as an example, its steel structure is co–laminated with Cu to reduce the surface impedance at its operating temperatures of 5–20 K [9]. The flat geometry makes REBCO–CC a feasible alternative coating material for BS to potentially reduce the resistive-wall impedance at operating temperatures of ~50 K [10]. Decreasing the impedance reduces the beam instabilities [11] and heat load to the BS [12].

Electron multipacting can occur inside the BS when certain resonant operating conditions are met [13]. The electric field wake created by passing bunches repeatedly accelerates electrons and results in avalanche–like electron multiplication and the formation of a stable and self–sustaining electron cloud (EC). The precondition for EC emergence in a BS, or another RF device, is a sufficiently high value of a secondary electron yield (SEY), defined as the number of secondary electrons leaving a surface after the incoming primary electron impact. If this ratio is above unity, the electrons can multiply, while a surface with a SEY below unity acts as an electron absorber, therefore, hindering EC build–up. The EC is a phenomenon commonly observed during the operation of storage rings [14–17] and RF devices [18,19] and is detrimental to the accelerator operation in many regards, such as EC–induced beam instabilities, leading to emittance growth [14], EC–induced heat–load deposited into the cryogenically cooled BS [12], and electron–induced gas desorption [20] that deteriorates the ultra–high vacuum (UHV) conditions in the BS. The EC–induced gas load is characterized by the electron–stimulated desorption (ESD) yield. The EC has a characteristic energy spectrum known from calculations, in situ measurements, and simulations [13,15,18,21,22], mostly consisting of low–energy electrons below 20 eV, and a smaller population of electrons accelerated by the electric field to hundreds of electron volts. Fortunately, both SEY and ESD yields decrease during electron irradiation to acceptable levels. This effect is leveraged in so–called beam–scrubbing [17,23], i.e., in situ electron and photon conditioning by circulating a beam when gradual decreases in EC activity and the dynamic vacuum effect are observed during the accelerator operation. These effects mainly pertain to accelerator storage rings, which aim to circulate bunched, positively charged beams at high intensity and high energy while maintaining a low emittance to generate high luminosity for physics research [24], such as the Future Circular Hadron Collider (FCC–hh) led by CERN [10] and the Super Proton–Proton Collider (SPPC) led by IHEP and CAS in China [25].

Previous studies [26–29] done on REBCO–CC samples (of the same type and provider studied here) produced experimental evidence for compatibility from the RF and superconducting perspectives across the necessary temperatures, fields, and current densities. This makes them potential candidates that are currently being investigated for BS construction to be used in future accelerators [27].

Hence, this manuscript investigates the compatibility of REBCO–CC surface properties linked to the dynamic vacuum effect and electron cloud at conditions representative of an accelerator application. This includes cryogenic temperatures and electron irradiation similar to that of an actual EC, i.e., with a focus on the entire 0–1 keV energy region. The following measurements use various electron beam–based methods to infer the sample's response to electron irradiation. First, the surface morphology and composition were investigated to verify that the REBCO coating was actually exposed and matched the manufacturer's specifications. The analytical methods we used were X-ray photoelectron spectroscopy (XPS) and scanning electron microscopy (SEM), conducted on a device equipped with an energy–dispersive X-ray spectroscopy (EDS) function. Second, the SEY was investigated as the main EC–driving property. SEY is a strongly surface–dependent property mostly derived from surface electronic properties that unwind from the chemical composition and surface state. Third, the ESD yield was measured as a function of energy and dose to estimate how the REBCO–CCs would react to an EC irradiation in an accelerator. The ESD yield is proportional to the amount of desorbable surface–bound contaminants that are normally present, even on UHV–grade cleaned surfaces.

## 2. Material and Methods

### 2.1. Samples Preparation

The two investigated samples of REBCO–coated conductors came from two different manufacturers, SuNAM and SuperOx. The samples were stored in a plastic box filled with air dehumidified by silica beads. Prior to measurements, the REBCO tapes were scissor–cut into 1 $cm^2$ large pieces, glued to flag–style copper sample holders with a silver–filled epoxy, and cured to 80 °C under a high vacuum.

The tapes have a layered structure deposited over a Hastelloy C276 used as a substrate, as visualized in Figure 1 on the right. There is a range of deposition methods [30], from which SuNAM uses RCE–DR (reactive co–evaporation by deposition and reaction) [31] and SuperOx uses PLD (pulsed laser deposition) [32]. Hence, the samples will be referred to as PLD–deposited and RCE–deposited to best represent the surface. The REBCO layer itself, i.e., the Gd–Ba–Cu–O, was deposited with 1600 nm and 900 nm thickness for RCE and PLD, respectively, and was grown on buffer underlayers [27]. Moreover, these studied samples were not the typical off–the–shelf REBCO tapes because they lacked the micrometer–sized protective overlayers typically present on industry-grade tapes. Indeed, this is done because the application in RF requires the HTS coating to be exposed, minding the small micrometer–sized skin depth. The RCE–deposited sample had its usual Ag overlayer removed by the manufacturer so effectively that it lacked Ag traces in the XPS spectrum, unlike the PLD–deposited sample. The protective micrometer–thick Ag overlayer had to be etched away from the PLD–deposited sample using a solution of $NH_3$:$H_2O_2$:$CH_3OH$ in 1:1:5 proportions, as detailed in [28], which inevitably affected its surface chemical state.

## 2.2. *Sample Analysis*

The SEM imaging, EDS analysis, and XPS analysis were acquired in this order after the SEY and ESD measurements were taken (in order to not interfere with the SEY and ESD results, which are both highly surface–sensitive).

### 2.2.1. Surface Morphology Imaging via SEM

Figure 2 presents the SEM images of the two different REBCO tape samples, PLD–deposited (left) and RCE–deposited (right). The apparent surface texture and morphology differences can be attributed to the different deposition methods. The high contrast of nanocrystals against the Gd–Ba–Cu–O substrate in the background is likely due to a combination of relatively higher atomic mass (higher Ba and Gd content) and a localized charging effect due to the dielectric nature of the oxide nanocrystals.

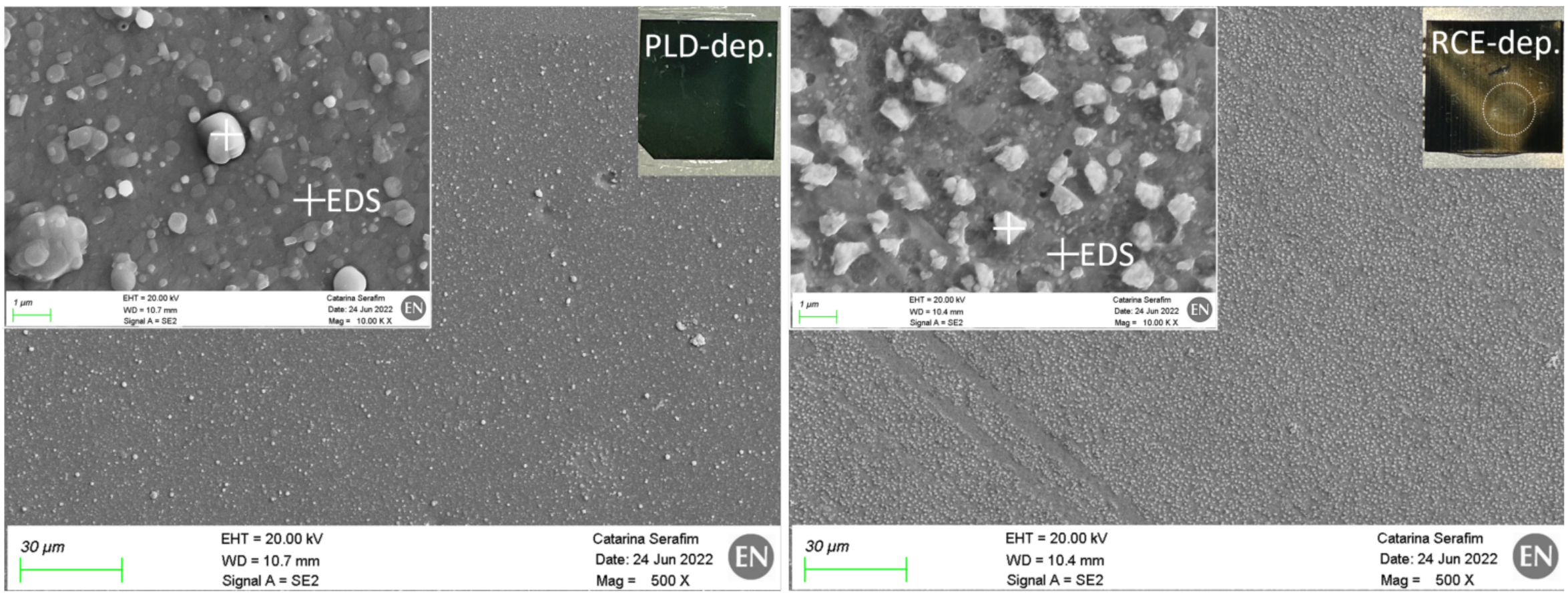

**Figure 2.** SEM images of REBCO samples with the well–visible nanocrystal–covered surface. Crosses represent the EDS measurement positions. **Left**: PLD–deposited, a wide range of crystal sizes. **Right**: RCE–deposited coating, uniformly sized crystals. Details in the text. The SEM images were taken on FE–SEM Zeiss Sigma.

The RCE coating process produces sharp uniformly sized and spaced nanocrystals, while the PLD coating process gives a wide range of oval–shaped nanocrystal sizes spaced at irregular intervals. These nanocrystals uniformly cover the entire surfaces of both samples.

### 2.2.2. Composition Analysis via EDS

The energy dispersive X-ray spectrometry analysis was performed on both studied samples to verify the surface chemical composition. EDS utilizes the X-ray spectrum emitted by a solid sample bombarded with a focused electron beam to obtain a localized chemical analysis. We identified the specific energy of the X-ray peaks characteristic for each element. The relative intensity of an X-ray line is approximately proportional to the mass concentration of an element. In a quantitative analysis, the concentrations of the elements present are determined by measuring the relative intensities for the lines identified in the spectrum for each element. The resulting concentrations of the identified elements are compiled in Table 1. However, the EDS analysis depth of ∼5 µm also incorporates all buffer layers and part of the Hastelloy substrate, which offsets the composition measurement.

For the RCE–deposited sample, the EDS spectrum was measured in an area with a nanocrystal and an area without, i.e., the substrate. The locations analyzed by EDS are marked by crosses in the insets of Figure 2. The precipitates segregated at the RCE–deposited surface may be Cu–O or Cu–Ba–O, resulting from the liquid process of RCE–DR to grow the Gd–Ba–Cu–O [31]. The PLD–deposited sample was analyzed in the same way, and the measured compositions derived from EDS spectra are listed in Table 1. Given the PLD growth method, the segregated precipitates at the surface are droplets of the same composition as the REBCO layer [32].

**Table 1.** Weight concentrations in percentage (%) of elements detected by the EDS method on the surface of PLD-deposited and RCE-deposited samples, at spots indicated in Figure 2. Analysis depth in ∼5 µm range.

| Sample | C | O | Cu | Ba | Gd |
|---|---|---|---|---|---|
| PLD–deposited: precipitate | 4.5 | 13.0 | 21.0 | 33.5 | 28.0 |
| PLD–deposited: Gd–Ba–Cu–O film | 5.7 | 13.2 | 21.0 | 33.3 | 26.8 |
| RCE–deposited: precipitate | 6.6 | 17.4 | 62.0 | 9.4 | 4.6 |
| RCE–deposited: Gd–Ba-Cu–O film | 4.4 | 13.1 | 19.1 | 29.2 | 34.2 |

2.2.3. Composition Analysis via XPS

The surface chemical analysis by XPS was performed in a room temperature UHV system equipped with a monochromated Al $K_\alpha$ X-ray source and a hemispherical electron energy analyzer collecting photoelectrons at normal emission angles. The energy scale was calibrated by setting the Cu $2p_{3/2}$ and the Au $4f_{7/2}$ states of sputter–cleaned metals to 932.6 and 84.0 eV, respectively. A mild charging effect was observed (deformed C 1s line) for the RCE–deposited sample. No charge compensation was applied. The atomic concentrations of the present elements were obtained assuming a uniform material and weighting the respective peak area of the XPS spectrum after background subtraction by the corresponding sensitivity factors. Table 2 lists the main elements found on the surface of the two REBCO samples. Elements such as Cl, P, F, Zn, and Ca were also detected, amounting to less than 10 atomic % in total. These elements likely come from post–production sample handling; for instance, Cl is introduced by epoxy–gluing. Meanwhile, the carbon, also observed in EDS spectra, is a common and dominant surface contaminant that has a naturally low SEY value if present in the sp2 state.

**Table 2.** Atomic concentrations in % of elements detected on the surface of as–received PLD–deposited and RCE–deposited samples by the XPS method. Analysis depth in the ∼nm range.

| Sample | C | O | Cu | Ba | Gd | N | Ag | Si |
|---|---|---|---|---|---|---|---|---|
| PLD–deposited: | 33.2 | 42.7 | 2.8 | 8.0 | 0.9 | 2.1 | 1.2 | 0.0 |
| RCE–deposited: | 49.3 | 36.3 | 1.7 | 1.9 | 0.4 | 1.4 | 0.0 | 5.3 |

The analysis depth in the nanometer range renders the XPS highly sensitive to the oxidized surface layer, making it challenging to infer the bulk REBCO composition. Still, the XPS surface chemical composition analysis, such as the EDS, could identify the main chemical constituents of the Gd-Ba-Cu-O coating. The chemical states of the different elements are known to greatly influence the SEY of a surface [33,34]. This is particularly the case for barium, which shows a maximum SEY below unity for the pure metallic state, while BaO displays a maximum SEY beyond 2 [33]. From the XPS spectra, the position of the Ba $3d_{5/2}$ core level at a binding energy of 780.1 eV for the PLD–deposited sample is compatible with the presence of BaO oxide [35]. The additional shift of the Ba $3d_{5/2}$ line to 780.4 eV for the RCE–deposited sample can be explained by a mild charging effect. Hence, in the presence of BaO with intrinsically high SEY, it is likely the elevated carbon surface concentration on the RCE–deposited sample that is responsible for the low SEY reported below.

### *2.3. Experimental Techniques*

#### 2.3.1. Experimental Setup

A newly developed analytical system was used to acquire the SEY and ESD data. The corresponding methodology for cryogenic measurements was described in detail in preceding articles [36,37].

The studied sample was mounted on a cryogenic manipulator installed in a baked UHV $\mu$–metal chamber held at room temperature, with a base pressure in the $10^{-10}$ mbar range. A Kimball ELG–2 electron gun irradiated the sample at normal incidence with a focused beam of 0–1.4 keV electrons at sub–nA to a few µA beam currents and a 2–3 mm wide flat–top beam profile. A custom–designed collector held at 300 K formed closed geometry over the studied cold sample and captured the emitted secondary electrons and neutral gas species.

The sample was biased to $-28$ V to impose a retarding potential that slowed the primary electron beam down to 0 eV. By doing so, both SEY and ESD energy dependence measurements started from 0 eV, as referenced in the sample vacuum level. Conversely, the electron conditioning was performed at a +46 V sample bias to minimize the gas desorption caused by the secondary electrons. Indeed, the electron energy was always corrected for the used bias.

#### 2.3.2. SEY Measurement

The SEY, denoted as $\delta$, is defined as the ratio between the primary electrons impinging the sample and the secondary electrons leaving the sample, irrespective of their physical origins. The collector current $I_C$ and the sample drain current $I_S$ were directly measured in our experiment. The beam current was recollected as a sum of the sample drain current and the collector current $I_B = I_S + I_C$. Then, the SEY was directly calculated from the observables.

$$\delta = \frac{I_{Secondary}}{I_{Primary}} = \frac{I_C}{I_B} = \frac{I_C}{I_C + I_S} \tag{1}$$

The SEY and ESD energy dependencies were measured point–wise, starting from 0 eV. The SEY data points were spaced 0.25 eV apart at low energy and up to 50 eV apart around 1.4 keV. The beam current of ∼0.5 nA preserved a signal–to–noise ratio (SNR) of ∼50 on the measured currents, while avoiding a significant conditioning effect during the measurement.

#### 2.3.3. ESD Measurement

The same collector also captures the electro-desorbed gas and directs it toward a calibrated quadrupolar mass spectrometer (QMS). The QMS absolute sensitivity $k_j$ [$A_j/mbar_j$] is known for each gas species $j$ of interest by an in situ calibration, so the partial pressure can be measured as $\Delta p_j = k_j \cdot \Delta i_j$. Knowing the pressure differential $\Delta p_j$ and the geometrically fixed collector conductance $C_j$ allows measuring the electro-desorbing gas flux of each monitored species. Finally, the ESD yield [molecule/$e^-$–] is given by the ratio of gas molecules, leaving the sample to the impinging primary electrons.

$$\eta_j = \frac{C_j \cdot \Delta p_j}{k_B \cdot T} \bigg/ \frac{I_B}{q_e} \tag{2}$$

In an ESD energy dependence measurement, the data points are spaced at 1 eV in the low–energy region and up to 300 eV around 0–1 keV. The beam current of ∼2 µA is sufficient to create a measurable pressure rise on the QMS, but low enough not to significantly damage the sample, as discussed in [36].

### *2.4. Measurement Procedure*

The SEY and ESD analyses were performed first, as they were the most sensitive to the sample surface state. All SEY and ESD measurements were taken at cryogenic temperatures below 20 K and were done on intact as–received spots of the same REBCO sample. An XPS analysis subsequently followed the SEY and ESD measurements to investigate the surface chemical composition of the studied samples. An intact spot was chosen to capture the as–received surface state. Finally, the SEM and EDS analyses were done to study the sample surface morphology and localize in–depth bulk chemical composition. The XPS, SEM, and EDS were performed ex situ at ambient temperatures, as they were temperature–independent in this case.

The two REBCO samples mounted on a flag–type sample holder were inserted into the UHV system via a load lock. Once the system was cold, the sample was rapidly heated to ∼100 K to desorb gases that could cryosorb from the background during the cool–down. Care was taken during the second and final cool–down from 100 K back to 12 K to ensure that the sample cooled down as the last, once again, mitigating the possibility of residual gas cryosorption. The sample was then ready for data–taking after the final cool–down when the system stabilized around $T_S$ = 12 K. Multiple SEY curves were acquired at different locations on the as–received sample to study variability within the sample itself. Measuring the SEY at the same locations at ambient temperatures prior to cool–down showed no tangible difference, as discussed below. Following the non–destructive SEY measurements, the ESD energy and dose dependencies were measured on two separate as–received spots of the same sample. The ESD energy dependence was measured first and was followed by a dose–dependent measurement at 300 eV to acquire a so–called conditioning curve. The SEY and ESD energy scans took about 5 min to finish, and the electron conditioning measurement took about 90 min. Once the ESD conditioning measurement was done, the SEY curve was acquired once again to capture the conditioned sample surface.

## **3. Results and Discussion**

### *3.1. Secondary Electron Yield*

The secondary electron yield measurements reported in Figure 3 provide a few important findings. First, there are no charging effects, even at ambient temperatures, despite the thick REBCO layer. This is essential since charging effects (often on dielectric surfaces) are detrimental in many technical applications, including applications in beam screens or RF devices.

Secondly, similar to technical–grade metals, the SEY of a REBCO-CC sample decreases under 300 eV of electron irradiation, a phenomenon known as electron conditioning. As most UHV–cleaned metal surfaces exhibit high SEY in their as–received air–exposed states [36,37], they typically condition during electron irradiation to reach SEY values at around 1, which are low enough to prevent electron multipacting.

The first stage of the SEY decrease is dominated by the ESD process, while graphitization of carbon-containing surface–bound contaminants dominates in the later stages, as proven independently by Cimino [38], Nishiwaki [39], and Scheuerlein [40], who linked the SEY decrease to graphitization. The conditioning process is mostly done at a dose of a few mC·mm$^{-2}$ and further electron irradiation at 300 eV does not lead to a significant SEY decrease. The SEY of the RCE–deposited sample conditions to an acceptable value of $\delta_{max}$ = 1.2 at 1 mC·mm$^{-2}$ dose. Conversely, the PLD-deposited sample only conditions to around $\delta_{max}$ = 2. The XPS analysis reveals two factors that explain the significant difference in SEY. Firstly, barium, which is present in BaO oxide form, is known to have a high SEY. Secondly, carbon contamination, when graphitized under electron irradiation, can reduce the SEY of surfaces close to unity. The RCE–deposited sample showed a higher atomic concentration ratio of carbon to barium than the PLD–deposited sample. This may partially account for the lower SEY observed after irradiation of the RCE–deposited sample. However, even after conditioning, the SEY of the PLD–deposited sample is too high for technical applications, and further SEY–reducing treatment is necessary to prevent multipacting.

Puig and Krkotić et al. [26] and Leveratto et al. [41] also reported high as–received SEY values of $\delta_{max} \approx 2.9$, as measured at ambient temperatures.

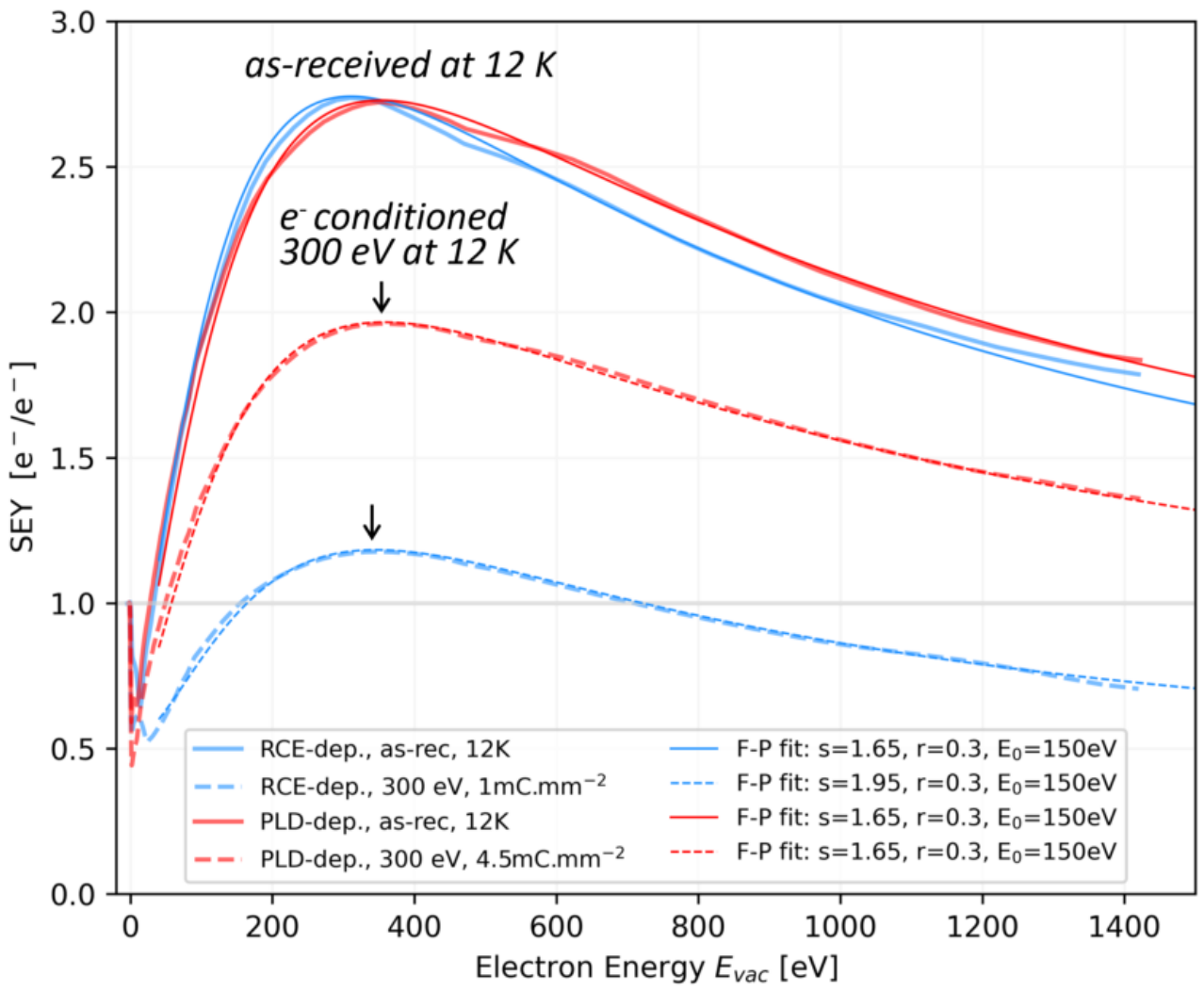

**Figure 3.** SEY energy dependence of REBCO samples held at 12 K measured in as–received and electron–conditioned states using 300 eV electrons at a few mC·mm$^{-2}$ doses, as indicated in the legend. The measured data points (not shown) are smoothed with a Savitzky–Golay filter [42] and plotted as a thick line, while the Furman–Pivi approximation [43] is the thin line.

The shapes of the SEY curves also closely resemble classical technical metals, as described by the $E_{max}$ and $\delta_{max}$ values. This is somewhat unexpected given the surface roughness observed in the SEM images that could point toward a more flat energy dependence that is typical for rough surfaces. The energy of maximum SEY $\delta_{max}$, the $E_{max}$, is important for the following reason. The typical energy spectrum of an electron cloud developed in an RF device consists of a high–energy peak at hundreds to thousands of eV [15,18,21,22]. The multipacting conditions can be better mitigated if the high–energy peak of EC is offset from the SEY peak position $E_{max}$, as is visible in the simulations done in [21].

Similar to the other technical–grade metals characterized in this setup, such as Cu, Al, and steel, the SEY of the studied REBCO coatings is temperature–invariant within the experimental precision and conditions in the investigated 12–300 K temperature and 0–1.4 keV energy ranges. Hence, the secondary electron emission seems to be agnostic to cryogenic temperatures, including around the superconducting transition around 90 K. The tentative explanation for the SEY stability in HTS under electron irradiation lies in the fact that electronic excitations have an energy range of units to tens of eV, while the binding energy of superconducting pairs is on the order of tens of meV, as theoretically calculated for HTS cuprates [44]. Recent measurements conducted at cryogenic temperatures [28] and further reasoning in [45] show that high–energy synchrotron radiation does not impair the superconducting properties in any way. This suggests that electron irradiation, which also decays into electronic excitations, is unlikely to impair the HTS performance. Therefore, the SEY can also be expected to remain unaffected by the conducting state of the HTS.

The experimental SEY curves can be fitted using a multi–component semi–empirical model, given in Equation (3), as implemented in PyECLOUD code [46] to facilitate further application. The three main components to fit the entire SEY curve, marked here as $\delta_{TOT}$, are the following: The true secondary electrons (TSE) part $\delta_{TSE}$ that dominate at primary energies above ∼20 eV and are fitted with a classical Furman–Pivi fit [43]. The elastically backscattered (EBS) part $\delta_{EBS}$ that dominates at sub $\sim$20 eV energies is approximated by a fit reportedly derived by Blaskiewicz [47]. Meanwhile, the inelastic backscattered (IBS) part $\delta_{IBS}$ starts at 0, but swiftly increases to the $r$ value, reaching 63% at 10 eV. An exponential function was chosen here for simplicity, but the low–energy SEY region generally deserves greater care given its crucial role in EC build–up.

$$\begin{aligned} \delta_{TOT} &= \delta_{TSE} + \delta_{EBS} + \delta_{IBS} \\ \delta_{TSE} &= (\delta_{max} - r_{BSE}) \cdot \frac{s(E/E_{max})}{s - 1 + (E/E_{max})^s} \\ \delta_{EBS} &= r_0 \cdot \left( \frac{\sqrt{E_0} - \sqrt{E_0 + E}}{\sqrt{E_0} + \sqrt{E_0 + E}} \right)^2 \\ \delta_{IBS} &= r_{BSE} \cdot (1 - e^{-E/10}) \end{aligned} \tag{3}$$

The parameters that best approximate the measured data are listed in the legend of Figure 3 and correspond to those suitable for technical metals. The following parameters were used here to accurately represent the measured SEY curves: reflectivity at 0 eV $r_0$ = 0.75, EBS cutoff at $E_0$ = 150 eV, skewness parameter $s$ = 1.65, and IBS fraction $r$ = 0.3. Lastly, the values of $\delta_{max}$ and $E_{max}$ of the fitted curve have to match the measured values.

### 3.2. *Electron Stimulated Desorption*

#### 3.2.1. Energy Dependence

ESD measurements were performed on both investigated samples at 12 K in an intact as–received state. Similarly to the SEY, the ESD curves acquired here for the two REBCO–CCs, displayed in Figure 4, also resemble regular metal surfaces. This applies to comparing the electro-desorbed gas composition, desorption thresholds, and nominal ESD yields and their maximum to technical–grade UHV–cleaned metals, such as Cu previously studied on this cryogenic setup [36,37] and other technical metals studied by other authors at ambient temperatures [48–53]. The energy thresholds for desorption are in the typical 5–10 eV region characteristic of technical–grade metals, such as Cu measured on the same setup [36], or extrapolated by Billard et al. [50]. The energy $E_{max}$ of the peak desorption yield $\eta_{max}$ of both samples lies in the 300–500 eV region. $H_2$ has the maximum yield and is followed by CO, $CO_2$, $CH_4$, and other UHV–typical (but less abundant) species. A sample storage more suitable for UHV components would likely lead to significantly lower ESD yields but similar energy and dose dependencies.

A semi-empirical fit was used to approximate the acquired experimental data for ESD yields and energy dependencies. The fit consisted of a log-normal distribution with an energy offset and cutoff at the desorption energy threshold $E_{thr}$, as described by the following equation.

$$\eta(E) = \eta_{max} \cdot \exp\left( \frac{-\ln^2((E - E_{thr})/E_{max})}{2\sigma^2} \right) \quad \text{, for} \quad E \geq E_{thr} \tag{4}$$

The peak ESD value $\eta_{max}$ at energy $E_{max}$ was used to normalize the energy distribution, whilst the $\sigma$ parameter determined its width. The legend in Figure 4 lists (for all measured gases) the relevant parameters necessary to reconstruct the energy dependence in the studied 0.1 keV region: $E_{thr}$, $\eta_{max}$, $E_{max}$, and $\sigma$. In practice, a constant was added to the parametric fit to model the non–zero background, which represents the detection limit of the used experimental setup.

The uncertainties were evaluated for $\eta_{max}$, as well as $E_{max}$ values, and are plotted in Figure 4 at the ESD yield peak value $\eta_{max}$. The bars represent a combined uncertainty of $\sim$30% at 1 $\sigma$ confidence interval, corresponding to 68% probability of containing the true value.

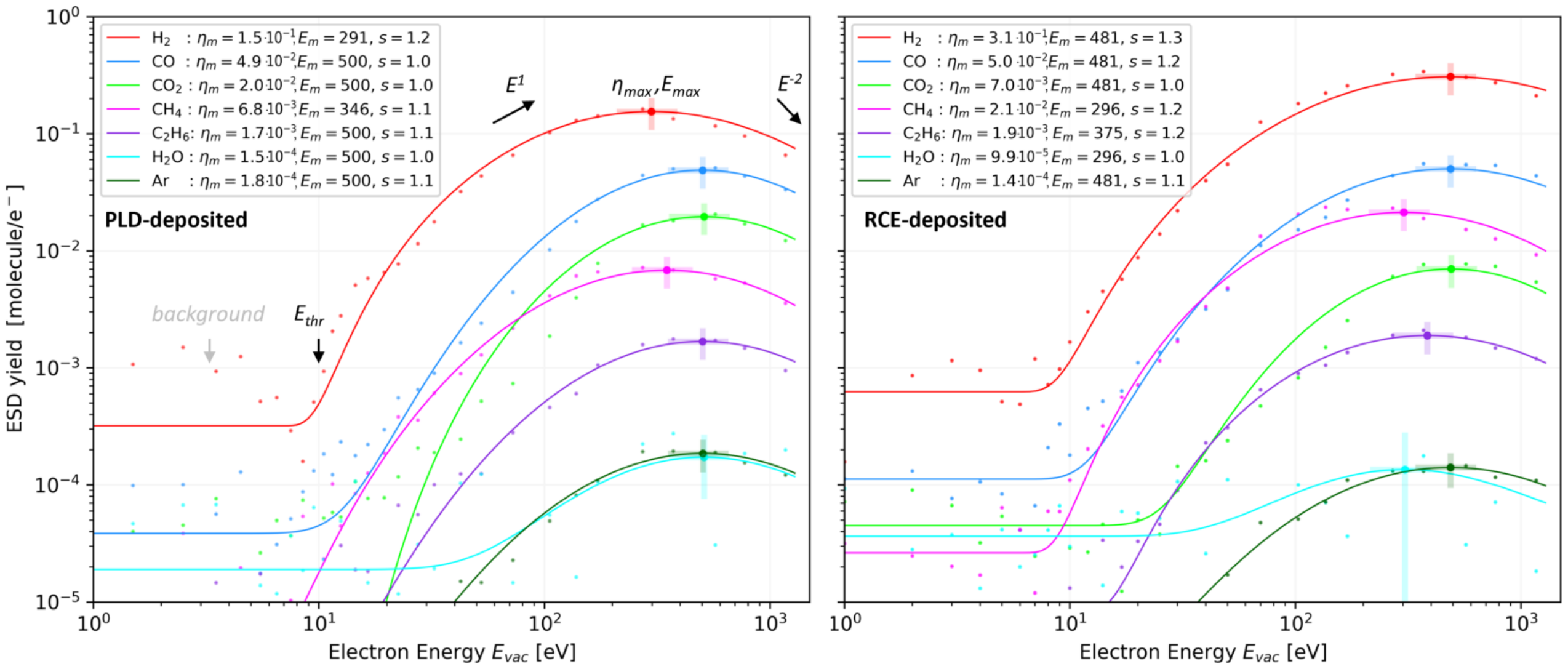

**Figure 4.** ESD energy dependence of as–received REBCO samples. The RCE–deposited sample (**right**) is characterized at 12 K and PLD–deposited at 19 K (**left**). Note the high uncertainty for water on the right plot, partially due to the high background and partially due to the low signal–to–noise ratio. The best–fit parameters for Equation (4) are listed in the legend. More details are in the text.

### 3.2.2. Dose Dependence

The conditioning curves, i.e., the ESD yield decreases with the electron doses, were acquired during a 300 eV electron irradiation performed at 12 K and are reported in Figure 5. The initial yields $\eta_0$, although slightly higher for the RCE–deposited sample, are characteristic of UHV–cleaned metal surfaces. The conditioning rates are just as rapid as for bare technical–grade metal surfaces and represent gas desorption from a non–porous surface. The slope of 0.3∼0.4 is typical for water electrodesorption from technical–grade metals. It is most likely the reason for a dynamic balance between electron stimulated water desorption and electron irradiation–induced synthesis. This is commonly observed in water electrodesorption because the $H_2O$ molecule is known to act as a chemical attractor [54] due to its high binding energy.

The irradiated RCE–deposited surface showed discoloration following the 300 eV electron conditioning at 12 K, as seen in the inset of Figure 2. Going from a dark to a light shade is unusual and the opposite of what happens to metals due to electron–induced surface graphitization. Instead, this bleaching effect could indicate a reduction of surface oxides and hydroxides.

The ESD conditioning curves measured here for the studied REBCO coatings also reveal no significant departure from the expected behavior when compared to the technical–grade Cu, Al, and steel previously measured using the same methods [36,37]. The data equally well compare to data from other authors who conditioned metals held at ambient temperatures in terms of the electro–desorbed gas composition, initial ESD yields $\eta_0$, inflection point location $D_1$, and conditioning rates $\alpha$. This includes work by Gomez–Goñi and Mathewson [52], Billard et al. [50], Achard [53] from CERN, Suzuki et al. [51], Kennedy [48], and Malyshev [49].

The fact that both the SEY and ESD decrease during electron irradiation has severe implications for applicability. As outlined in Section 1, the electron conditioning is leveraged in accelerators during beam–scrubbing [17,23] to bring down the EC activity by decreasing the SEY and mitigating the dynamic vacuum effect by conditioning the desorption yields.

An empirical fit, initially proposed by Malyshev [49] for photodesorption, was used here to approximate the ESD yield dose dependence and possibly extrapolate to higher electron doses. The fit builds on a classical power law whose exponent $\alpha$ controls the slope in the log–log plot at high doses $D \geq 0.1\,\mathrm{mC{\cdot}mm^{-2}}$. Parameters $D_0$ and $D_1$ are added to the fraction to extend the applicability toward low doses in a way that the curve asymptotically approaches the constant initial ESD yield $\eta_0$ as $D \to 0$. The parameter $D_0$ represents the dose imparted at the lowest measurable data point (here, it is the first displayed data point). Parameter $D_1$ is solely used to position the end of the initial plateau.

$$\eta(D) = \eta_0 \cdot \left( \frac{D + D_1}{D_0 + D_1} \right)^{-\alpha} \tag{5}$$

Values for $\eta_0$, $D_1$, and $\alpha$ parameters in Equation (5) can be read directly from the legends in Figure 5 and plugged into the presented parametric fits for the ESD yield dose dependence. The same uncertainty analysis leading to $\sim$30 % at 1 $\sigma$ confidence interval applies to the ESD yield values presented here too, although not plotted.

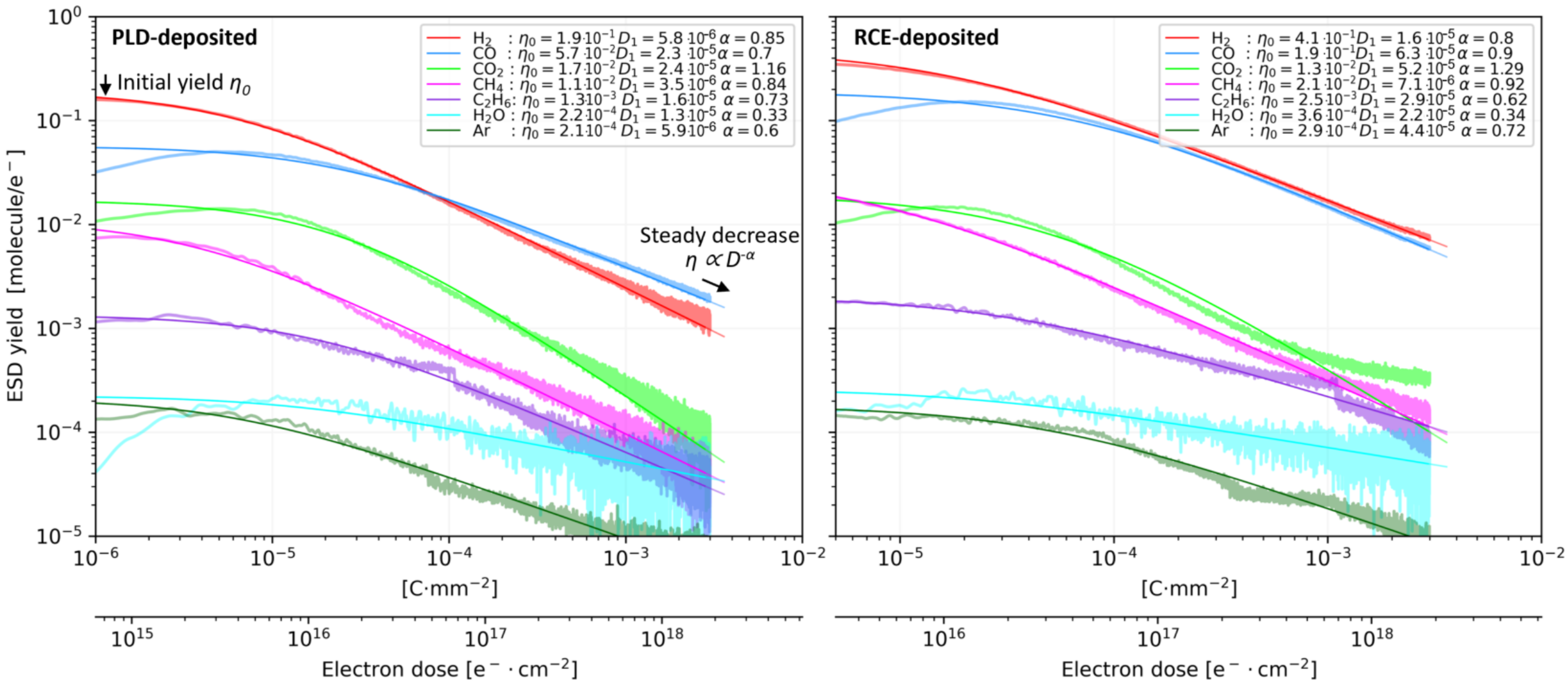

**Figure 5.** ESD dose dependence measured at 300 eV for PLD–deposited (**left**) and RCE–deposited (**right**) REBCO samples held at 12 K starting from an as–received state. Best–fit parameters for Equation (5) from [49] is listed in the legend. More details are in the text.

## 4. Summary and Conclusions

This electron beam–based laboratory investigation of REBCO–CCs aims to characterize surface properties related to electron clouds in accelerators, including the secondary electron yield (SEY), which is a determining factor for electron multipacting, and electron stimulated desorption (ESD), which links dynamic pressure rise to EC activity during operation. The presented results are applicable to other types of RF devices that use HTS, such as radio–frequency waveguides and cavities, which may also suffer from electron cloud or multipacting effects.

We first identified the surface morphology and composition using SEM imaging with an EDS capability. SEM imaging showed a flat surface structure covered with segregated nanocrystalline precipitates with characteristic sizes in the micrometer range. The ESD conditioning rate and the shape of measured SEY curves generally resemble those of bare technical–grade metals, which supports the SEM observation of a flat, non–porous, surface structure covered with micrometer–sized precipitates. The EDS measurement is bulk–sensitive at the micrometer range, and considering the characteristic micrometer–sized skin depths, the EDS results are relevant to the HTS properties. Meanwhile, the XPS measurement with a probe–depth in the nanometer range makes it highly surface sensitive, rendering the XPS results relevant to the SEY properties. Both EDS and XPS analyses could identify the main chemical constituents at the surface of the Gd–Ba–Cu–O film.

Knowing the morphology and composition of the samples and that the REBCO was directly exposed on the surface, we held the samples at 12 K and studied their response to electron irradiation. The energy range of 0–1.4 keV was chosen to cover the entire relevant EC spectrum [13,15,21]. Meanwhile, 300 eV irradiation was selected for electron conditioning studies, as it represents the high–energy part of the EC, which is responsible for the conditioning effect [38], i.e., the decrease of SEY and ESD yields during electron irradiation commonly observed during beam–scrubbing of an accelerator [20,23]. We measured the SEY energy dependence in the 0–1.4 keV range both before and after the 300 eV irradiation with a clear conclusion that the SEY decreases in the entire studied range. The SEY of the RCE–deposited sample decreased to a value of $\delta_{max}$ = 1.2 at 1 mC·mm$^{-2}$ electron dose, which is a value acceptable for many applications. Conversely, the PLD–deposited sample SEY only conditioned to $\delta_{max} \approx 2$ at a 4.5 mC·mm$^{-2}$ dose. The reason for this is unclear, but based on the XPS analysis, it can be hypothesized that the etched PLD–deposited sample lacks surface–bound carbon contaminants that would graphitize into a sufficiently thick graphitic overlayer with intrinsically low SEY values. The XPS spectrum seconds this hypothesis since the RCE–deposited sample had 49% carbon content whilst the PLD–deposited sample only had 33%, resulting in a smaller SEY reduction.

The electron conditioning was studied up to a few mC·mm$^{-2}$, which is an electron dose rapidly achievable during an accelerator operation. The RCE-deposited case represents a proof-of-concept capability of REBCO-type coatings to electron condition to SEY values sufficiently low not to trigger electron multipacting, resulting in EC build-up in the beam screen. Parallel studies by Puig et al. [26] also consider depositing a thin, anti-multipacting, low-SEY layer of amorphous carbon coating atop the REBCO-CCs to alleviate the need for beam scrubbing. Therefore, future SEY and ESD studies could focus on characterizing carbon-coated REBCO-CCs at cryogenic conditions.

The acquired ESD data represent a link between the EC and the dynamic vacuum effect caused by EC electron-desorbing gases from the cryogenic BS wall. Hence, the ESD yield measurements also covered the energy and dose dependencies similar to the SEY. Irradiation with 300 eV electrons at a dose of a few mC·mm$^{-2}$ resulted in a decrease in all measured ESD yields by about 2 orders of magnitude for both studied REBCO-CCs held at 12 K. The ESD yields were also measured for the two samples as functions of energy in the 0–1 keV range. The acquired ESD results are similar to the usual technical-grade UHV-cleaned metals previously studied in this setup [36,37] and by other authors under similar conditions [48–53], in terms of desorption thresholds, nominal ESD yields, and conditioning rates of all UHV–typical gases. Hence, the dynamics of the beam–induced gas desorption from REBCO–CCs can be expected to be very similar to the technical-grade metals currently used as the baseline [55]. Moreover, the ESD yield reported here can be considered as a proxy for photon– and ion–stimulated desorption yields that are known to correlate with the ESD [36,56–58].

**Author Contributions:** M.H., V.B. and S.C. conceived and designed the study. V.B. and B.H. designed the SEY and ESD measurement apparatus. M.H. commissioned the apparatus, performed the SEY and ESD measurements, and analyzed the data. C.S. and V.P. conducted the XPS and SEM and EDS analyses and interpreted the data. P.K., S.C., T.P. and J.G. provided valuable feedback on REBCO superconductivity and edited the manuscript. M.H., P.K., V.B. and S.C. interpreted the results. A.R. and M.H. prepared the samples. M.H. wrote and edited the manuscript with contributions from all authors. All authors have read and agreed to the published version of the manuscript.

**Funding:** The authors acknowledge the support from the HL–LHC project and CERN funding FCCGOV–CC-0208 (KE4947/ATS). The authors acknowledge the support and samples provided by Superox and SuNAM. M. Haubner acknowledges partial support from the Czech Technical University in Prague, grant number: SGS21/149/OHK2/3T/12. A. Romanov acknowledges MSCA-COFUND-2016-754397 for the PhD grant. A. Romanov, T. Puig, J. Gutierrez acknowledge funding from PID2021-127297OB-C21 and CEX2019-000917-S.

**Data Availability Statement:** Data are available upon reasonable request.

**Acknowledgments:** The authors would like to acknowledge the support and samples provided by SuperOx and SuNAM. The authors acknowledge the support from the EN-MME group for providing the tools for SEM and EDS analysis.

**Conflicts of Interest:** The authors declare no conflict of interest.

## Abbreviations

The following abbreviations are used in this manuscript:

| | |
|---|---|
| BS | Beam Screen |
| CC | Coated Conductor |
| EC | Electron Cloud |
| EDS | Energy Dispersive X-ray Spectroscopy |
| ESD | Electron Stimulated Desorption |
| FE-SEM | Field-Emission SEM |
| HTS | High Temperature Superconductor |
| LTS | Low Temperature Superconductor |
| LHC | Large Hadron Collider |
| PLD | Pulsed Laser Deposition |
| QMS | Quadrupolar Mass Spectrometer |
| RCE-DR | Reactive Co-Evaporation by Deposition and Reaction |
| REBCO | Rare-Earth Barium Copper Oxide |
| RF | Radio Frequency |
| SEM | Secondary Electron Microscope |
| SEY | Secondary Electron Yield |
| SNR | Signal to Noise Ratio |
| UHV | Ultra-High Vacuum |
| XPS | X-ray Photoelectron Spectroscopy |